# Necessity of selfconsistent calculations for the electromagnetic field in probing the nuclear symmetry energy using pion observables in heavy-ion collisions

Gao-Feng Wei,[1,2,][*] Chang Liu,[1] Xin-Wei Cao,[3] Qi-Jun Zhi,[1,2]
Wen-Jun Xiao,[1] Chao-Yun Long,[4] and Zheng-Wen Long[4]

[1]*School of Physics and Electronic Science, Guizhou Normal University, Guiyang 550025, China*
[2]*Guizhou Provincial Key Laboratory of Radio Astronomy and Data Processing,
Guizhou Normal University, Guiyang 550025, China*
[3]*School of Mechanical and Material Engineering,
Xi'an University of Arts and Sciences, Xi'an 710065, China*
[4]*College of Physics, Guizhou University, Guiyang 550025, China*

Within an isospin- and momentum-dependent transport model, we investigate the necessity of selfconsistent calculations for the electromagnetic field in probing the nuclear symmetry energy using pion observables in heavy-ion collisions at intermediate energies. To this end, we perform the $^{96}$Ru + $^{96}$Ru collisions at 400 MeV/nucleon with two calculations scenarios for the electromagnetic field including the selfconsistent calculation and the most used Liénard-Wiechert formula, while the latter is a simplified one of the complete Liénard-Wiechert formula by neglecting the radiation field for practical calculations in heavy-ion collisions at intermediate and/or relativistic energies. As a comparison, we also consider the static Coulomb field formula for calculations of the electromagnetic field in heavy-ion collisions. It is shown that the most used simplified Liénard-Wiechert formula is not enough for the electromagnetic field calculation because the absent radiation field in this formula also affects significantly the charged pions as well as their $\pi^-/\pi^+$ ratio. Moreover, we also examine effects of the electromagnetic field in these scenarios on the double $\pi^-/\pi^+$ ratio of two isobar reaction systems of $^{96}$Ru + $^{96}$Ru and $^{96}$Zr + $^{96}$Zr at 400 MeV/nucleon. It is shown that the double $\pi^-/\pi^+$ ratio of two reactions tends to be less affected by the electromagnetic field calculation scenario and thus can still be an effective probe of the nuclear symmetry energy in heavy-ion collisions. Therefore, according to these findings, it is suggested that the selfconsistent calculation for the electromagnetic field should be carefully taken into account when using the pion observables to probe the nuclear symmetry energy in heavy-ion collisions.

## I. INTRODUCTION

As one of central issues in nuclear physics, the equation of state (EoS) of asymmetric nuclear matter (ANM) characterizes fundamental properties of the nuclear medium both in nuclei and heavy-ion collisions (HICs) as well as in many astrophysical objects, see, e.g., Refs. [1–9] for comprehensive reviews. Presently, on the isospin-independent term of EoS of ANM, i.e., EoS of symmetric nuclear matter (SNM), significant progress has been achieved [10–14], it is however, knowledge on the isospin-dependent part of EoS of ANM, i.e., nuclear symmetry energy, is still unsatisfactory. Certainly, around and below the saturation density $\rho_0$, the nuclear symmetry energy has been relatively well determined from the empirical liquid-drop mass formula [15, 16] as well as the data of finite nuclei [17, 18]. Therefore, the main task in this issue is the determination of nuclear symmetry energy at high densities. In nature, terrestrial experiments and astrophysical observations are two main tools in constraining the nuclear symmetry energy at high densities. In terrestrial laboratories, HICs indeed can be used to produce the high-density nuclear matter, but necessary comparisons with the corresponding theoretical simulations usually lead to either supersoft [19, 20] or superstiff [21] as well as moderately soft predictions [22–24] for the nuclear symmetry energy at high densities, strongly depending on the used models and/or data [24–26]. Certainly, using the observed gravitational wave signal GW170817 [27, 28] as well as the millisecond pulsar PSR J0740+6620 with mass $2.14^{+0.10}_{-0.09}M_{\rm sun}$ [29], some studies [30, 31] have claimed recently that the superstiff and supersoft high-density nuclear symmetry energy can already be excluded. Nevertheless, yet even so, constraints for the nuclear symmetry energy at high densities are still far from satisfactory compared with that around and/or below the saturation density. This is mainly because the isovector part of nuclear interactions is significantly weaker than the isoscalar part, and thus the isospin signals are usually interfered by other poorly known factors in theoretical simulations and experimental measurements. To this situation, a model comparison project [32–34] was carried out to quantify the model dependence as well as the uncertainties in theoretical simulations. Correspondingly, a symmetry energy measurement experiment, i.e., S$\pi$RIT project [35, 36], was also conducted at RIKEN-RIBF in Japan in the past few years. Along this direction, we shall investigate and eliminate in this article uncertainties of the electromagnetic field in HICs at intermediate energies.

Electromagnetic (EM) interaction plays an important role in the evolution of charged particles in HICs. To evaluate accurately effects of the EM field in HICs at intermediate and/or relativistic energies, one should in principle calculate selfconsistently the EM field from the Maxwell equations. In terms of the scalar potential $\varphi$ and vector potential $\mathbf{A}$ of EM fields, one can express the electric and magnetic fields as

$$\mathbf{E} = -\nabla\varphi - \frac{\partial \mathbf{A}}{\partial t}, \quad (1)$$

$$\mathbf{B} = \nabla \times \mathbf{A}, \quad (2)$$

and then incorporate them into transport equation such as the Boltzmann-Uehling-Uhlenbeck (BUU) type [37]. Obviously, the EM fields effects in this manner can be considered selfconsistently in HICs. Considering the fact that nucleons can be treated as point-like particles in HICs, one can arrive from Maxwell equations to the complete Liénard-Wiechert (LW) formulas [38, 39], i.e.,

$$e\mathbf{E}(\mathbf{r},t) = \frac{e^2}{4\pi\varepsilon_0}\sum_n Z_n \left\{ \frac{(c^2 - v_n^2)(c\mathbf{R}_n - R_n\mathbf{v}_n)}{(cR_n - \mathbf{R}_n \cdot \mathbf{v}_n)^3} \right. \\ \left. + \frac{\mathbf{R}_n \times \left[(c\mathbf{R}_n - R_n\mathbf{v}_n) \times \dot{\mathbf{v}}_n\right]}{(cR_n - \mathbf{R}_n \cdot \mathbf{v}_n)^3} \right\}, \quad (3)$$

$$e\mathbf{B}(\mathbf{r},t) = \frac{e^2}{4\pi\varepsilon_0 c}\sum_n Z_n \left\{ \frac{(c^2 - v_n^2)(\mathbf{v}_n \times \mathbf{R}_n)}{(cR_n - \mathbf{R}_n \cdot \mathbf{v}_n)^3} \right. \\ \left. + \frac{\mathbf{R}_n \times \left\{\mathbf{R}_n \times \left[(c\mathbf{R}_n - R_n\mathbf{v}_n) \times \dot{\mathbf{v}}_n\right]\right\}}{R_n(cR_n - \mathbf{R}_n \cdot \mathbf{v}_n)^3} \right\}, (4)$$

where $Z_n$ is the charge number of the $n$th particle at the position $\mathbf{r}_n$, and $\mathbf{R}_n = \mathbf{r} - \mathbf{r}_n$ is the relative position of the field point $\mathbf{r}$ to the source point $\mathbf{r}_n$. The summation runs over all charged particles with a velocity of $\mathbf{v}_n$ at the retarded time $t_n = t - |\mathbf{r} - \mathbf{r}_n|/c$. Obviously, the first term depends on the velocity $\mathbf{v}_n$ of the charged particle and thus is usually called the velocity fields, while the second term depends linearly on the acceleration $\dot{\mathbf{v}}_n$ of the charged particle and represents the radiation fields, i.e., acceleration fields. In principle, using Eqs. (3) and (4) also allows to the selfconsistent calculation for the EM fields in HICs. Nevertheless, since the full calculations of EM fields using Eqs. (3) and (4) are rather complicated, most of heavy-ion transport models usually employ the simplified Liénard-Wiechert (SLW) formula by neglecting the radiation fields in Eqs. (3) and (4) for practical calculations in HICs at intermediate [40–42] and/or relativistic energies [43, 44]. In fact, the effects of radiation/acceleration fields have never been estimated quantitatively in HICs at intermediate and/or relativistic energies. Moreover, it should be emphasized that the decay of acceleration fields (for the distance as $R_n^{-1}$) is an order of magnitude slower than the velocity fields (for the distance as $R_n^{-2}$) [38, 39]. To this situation, a natural question is whether the effects of acceleration fields are negligible in HICs, and answers to this question are undoubtedly of importance in probing the EoS of ANM at intermediate energies and/or in characterizing the chiral effects at relativistic energies.

---
[*] Corresponding author. E-mail: wei.gaofeng@gznu.edu.cn

The main purpose of this article is to examine and eliminate the uncertainties of electromagnetic fields calculations in HICs at intermediate energies, especially for the light mass charged pions and/or their ratios, since they are found to be very sensitive to the high-density nuclear symmetry energy [3, 4, 25, 45, 46] but still interfered by other incompletely known uncertainties [34, 47–50].

## II. THE MODEL

This study is carried out within an isospin- and momentum-dependent BUU (IBUU) transport model [37, 51, 52]. To estimate effects of the EM fields in HICs, two calculations scenarios for the EM fields produced in HICs are used. The first is the most used SLW formulas, i.e., Eqs. (3) and (4) under neglecting the acceleration fields (i.e., $\dot{\mathbf{v}}_n = 0$), and the second is the selfconsistent calculations using Eqs. (1) and (2), in which the scalar potential $\varphi$ and vector potential $\mathbf{A}$ are also calculated selfconsistently from sources of the charge $q_n = Z_n e$ and the current $\mathbf{j}_n = Z_n e \mathbf{v}_n$. As a comparison, we also include the results using the static Coulomb field formula (i.e., nonrelativistic limit of Eqs. (3) and (4)) for the calculations of EM fields in HICs. Certainly, as shown in Refs. [39–44, 53, 54], this scenario can already be excluded in HICs at intermediate and/or relativistic energies because the necessary retarded effects of relativistic motions are obviously absent.

While for the nuclear interaction used in this study, we use an isospin- and momentum-dependent interaction similar to our previous studies [53, 54], i.e.,

$$U(\rho, \delta, \vec{p}, \tau) = A_u(x)\frac{\rho_{-\tau}}{\rho_0} + A_l(x)\frac{\rho_\tau}{\rho_0} + \frac{B}{2}\left(\frac{2\rho_\tau}{\rho_0}\right)^\sigma(1-x) \\ + \frac{2B}{\sigma+1}\left(\frac{\rho}{\rho_0}\right)^\sigma(1+x)\frac{\rho_{-\tau}}{\rho}\left[1+(\sigma-1)\frac{\rho_\tau}{\rho}\right] \\ + \frac{2C_l}{\rho_0}\int d^3p' \frac{f_\tau(\vec{p'})}{1+(\vec{p}-\vec{p'})^2/\Lambda^2} \\ + \frac{2C_u}{\rho_0}\int d^3p' \frac{f_{-\tau}(\vec{p'})}{1+(\vec{p}-\vec{p'})^2/\Lambda^2}, \quad (5)$$

and the $x$-dependent parameters $A_l(x)$ and $A_u(x)$ are expressed in forms of

$$A_l(x) = A_{l0} - \frac{2B}{\sigma+1}\left[\frac{(1-x)}{4}\sigma(\sigma+1) - \frac{1+x}{2}\right], \quad (6)$$

$$A_u(x) = A_{u0} + \frac{2B}{\sigma+1}\left[\frac{(1-x)}{4}\sigma(\sigma+1) - \frac{1+x}{2}\right]. \quad (7)$$

Here, the parameter $x$ relevant to the spin(isospin)-dependent parameter $x_0$ via $x = (1+2x_0)/3$ in the density-dependent term of original Gogny effective interactions affects only the isovector properties of ANM at nonsaturation densities. It should be mentioned that

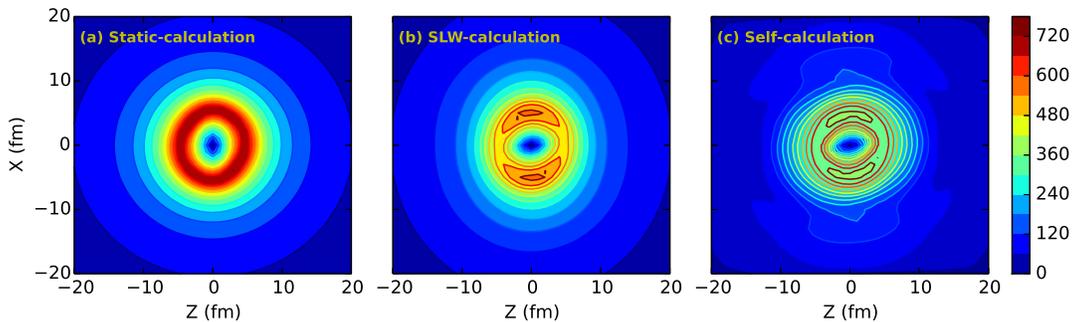

FIG. 1. (Color online) Contours of the electric fields $e|\mathbf{E}|$ (MeV$^2$) in the $X$-o-$Z$ reaction plane at the maximum compress stage in central $^{96}$Ru+$^{96}$Ru collisions at 400 MeV/nucleon with three different calcualtions scenarios.

in the present version of IBUU model we have updated the nuclear mean-field interaction through replacing the density-dependent term of original Gogny effective interaction [55], i.e.,

$$V_D = t_0(1 + x_0 P_\sigma)\big[\rho\big(\frac{\mathbf{r}_i + \mathbf{r}_j}{2}\big)\big]^\alpha \delta(\mathbf{r}_{ij}), \quad (8)$$

by a separate density-dependent scenario [56–58], i.e.,

$$V'_D = t_0(1 + x_0 P_\sigma)[\rho_{\tau_i}(\mathbf{r}_i) + \rho_{\tau_j}(\mathbf{r}_j)]^\alpha \delta(\mathbf{r}_{ij}), \quad (9)$$

since some studies related to nuclear structure have already shown that very satisfactory agreement with the corresponding experiments can be achieved using the separate density-dependent scenario such as the binding energies, single-particle energies, and electron scattering cross sections for $^{16}$O, $^{40}$Ca, $^{48}$Ca, $^{90}$Zr, and $^{208}$Pr [59, 60]. Moreover, to better fit the high momentum behaviors of nucleon optical potential extracted from nucleon-nucleus scattering experiments [61], we have readjusted the parameters $A_{l0}$, $A_{u0}$, $B$, $\sigma(\equiv \alpha + 1)$, $C_l$, $C_u$ and $\Lambda$ using empirical constraints on properties of nuclear matter at $\rho_0 = 0.16$ fm$^{-3}$, i.e., the binding energy $-16$ MeV, the incompressibility $K_0 = 230$ MeV for SNM, the isoscalar effective mass $m_s^* = 0.7m$, the isoscalar potential at infinitely large nucleon momentum $U_0^\infty(\rho_0) = 75$ MeV, and the symmetry potential at infinitely large nucleon momentum $U_{sym}^\infty(\rho_0) = -10$ MeV as well as the symmetry energy $E_{sym}(\rho_0) = 32.5$ MeV. The values of these parameters are $A_{l0} = -76.963$ MeV, $A_{u0} = -56.963$ MeV, $B = 141.963$ MeV, $C_l = -53.931$ MeV, $C_u = -106.257$ MeV, $\sigma = 1.2652$, and $\Lambda = 2.424 p_{f0}$, where $p_{f0}$ is the nucleon Fermi momentum in SNM at $\rho_0$. On the other hand, to improve the accuracies of theoretical simulations of HICs, we have also considered the pion potential and the $\Delta$ isovector potential in this study. Specifically, for the pionic momentum higher than 140 MeV/$c$, we use the pion potential based on $\Delta$−hole model of the form used in Ref. [62], and for the pionic momentum lower than 80 MeV/$c$, we adopt the pion potential of the form used in Refs. [63–65], while for the pionic momentum falling into the range from 80 to 140 MeV/$c$, we use an interpolative pion potential constructed by O. Buss in Ref. [62]. For the $\Delta$ potential, guided by the earlier studies [66, 67] and according to the decay mechanism of $\Delta$ resonances, we use an isospin-dependent $\Delta$ potential, i.e.,

$$U(\Delta^{++}) = f_\Delta U(p), \quad (10)$$

$$U(\Delta^+) = f_\Delta\big[\frac{1}{3}U(n) + \frac{2}{3}U(p)\big], \quad (11)$$

$$U(\Delta^0) = f_\Delta\big[\frac{2}{3}U(n) + \frac{1}{3}U(p)\big], \quad (12)$$

$$U(\Delta^-) = f_\Delta U(n), \quad (13)$$

where the factor $f_\Delta = 2/3$ is introduced to consider the fact that the depth of nucleon potential is approximately $-50$ MeV, while that of the $\Delta$ potential is empirically constrained around $-30$ MeV [62–65].

### III. RESULTS AND DISCUSSIONS

To help understand effects of the EM fields on the charged pions and their $\pi^-/\pi^+$ ratio, it is necessary to have a global picture of the EM fields produced in HICs. Considering that the magnetic field itself does not directly but indirectly through generating an induced electric field affects the energy of charged baryons and thus the multiplicities of total pions[1], we therefore focus on detecting the differences of the electrical fields in three calculations scenarios. Shown in Fig. 1 are the strength contours of the electric fields $|e\mathbf{E}|$ in the $X$-o-$Z$ reaction plane at the maximum compress stage in central $^{96}$Ru+$^{96}$Ru collisions at 400 MeV/nucleon with three calculations scenarios. First, compared with the electric field calculated using the static Coulomb field formula, the anisotropic characteristics of the electric field due to relativistic retarded effects are obvious either the SLW formula or the selfconsistent calculation scenario is used.

---

[1] As to the differential pions, such as rapidity and/or azimuth distribution of pions, the magnetic field still has direct influences on them.

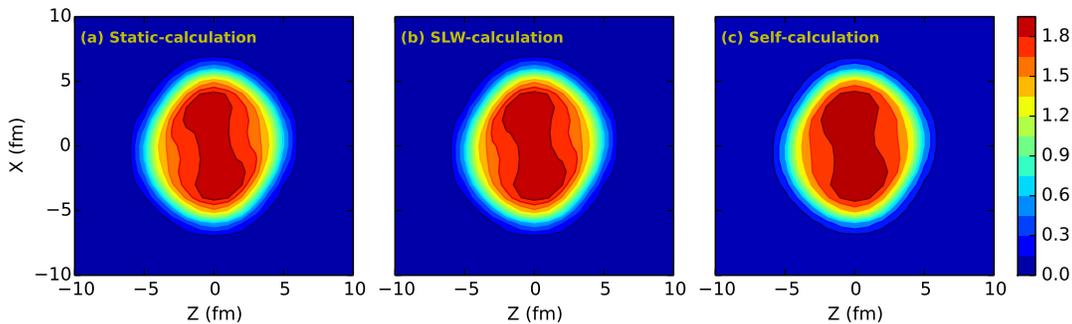

FIG. 2. (Color online) Contours of the densities $\rho/\rho_0$ in the $X$-o-$Z$ reaction plane at the maximum compress stage in central $^{96}$Ru+$^{96}$Ru collisions at 400 MeV/nucleon with three different calcualtions scenarios.

Second, the electric field with the selfconsistent calculations due to incorporating contributions of the radiation fields is overall weaker than that calculated using the SLW formula, while the latter is also weaker due to relativistic retarded effects than that calculated using static Coulomb field formula. This feature naturally will reduce the repulsive effects between protons especially in the $Z$ direction (i.e., beam direction) and thus get the densities of compress region (or the high-density region) more or less denser (larger). Indeed, as shown of the density contours in the $X$-o-$Z$ plane at the maximum compress stage in Fig. 2, we can see a slightly larger cover of the high-density region mainly in the $Z$ direction with the selfconsistent calculation scenario than that with the SLW formula, and the latter also covers a slightly larger area in the $Z$ direction compared to that with the static Coulomb field formula. Certainly, because the electric field force is much smaller than the nuclear force, we do not expect the global dynamics of nuclear reactions are changed by the EM fields calculations scenarios, this can also be confirmed by the similar density distributions in these three cases as shown in Fig. 2. Also, because of the symmetries in the plane perpendicular to the beam direction, consistent with our previous studies [53, 54] in studying the differences of electric fields between using the SLW formula and the static Coulomb field formula, the anisotropic feature for the electric fields is also invisible in the $X$-o-$Y$ plane even using the selfconsistent calculation scenario.

Now, we examine effects of the EM fields produced in HICs with three different calculations scenarios on the charged pions as well as their $\pi^-/\pi^+$ raitos. To compare with the effects of symmetry energy and/or potential on them, we also take several different values for the $x$ parameter that yields the symmetry energy from the stiff with $x = -1$ to the soft with $x = 1$. Shown in Fig. 3 are the final multiplicities of $\pi^-$ and $\pi^+$ generated in central $^{96}$Ru+$^{96}$Ru collisions at 400 MeV/nucleon with three calculations scenarios. First, consistent with previous findings in Ref. [68], the multiplicities of $\pi^-$ are more sensitive to the nuclear symmetry energy compared to that of $\pi^+$ since the $\pi^-$ mesons are mostly pro-duced from neutron-neutron inelastic collisions. Second, the multiplicities of both $\pi^-$ and $\pi^+$ are significantly larger with the selfconsistent calculation scenario than that with the SLW formula, while the latter are also significantly larger than that with the static Coulomb field formula. Moreover, these effects of electric fields on $\pi^+$ are approximately 2 times larger than that on $\pi^-$. Actually, as aforementioned, the weaker anisotropic electric fields and thus the larger cover of high-density region in $Z$ direction get some protons gaining larger kinetic energy and thus leading to more production of $\pi^+$ through the $p+p \to p+\pi^+$ inelastic channel. Certainly, the neutrons are not affected directly by the electric fields, it is however, secondary collisions between neutrons and energetic protons as well as neutrons coupling to charged $\Delta$ resonances through the $\Delta^+ \leftrightarrow n+\pi^+$ and $\Delta^- \leftrightarrow n+\pi^-$ channels can increase the kinetic energy of neutrons, and these energetic neutrons are responsible for the increased production of $\pi^-$ through the $n+n \to p+\pi^-$ inelastic channel.

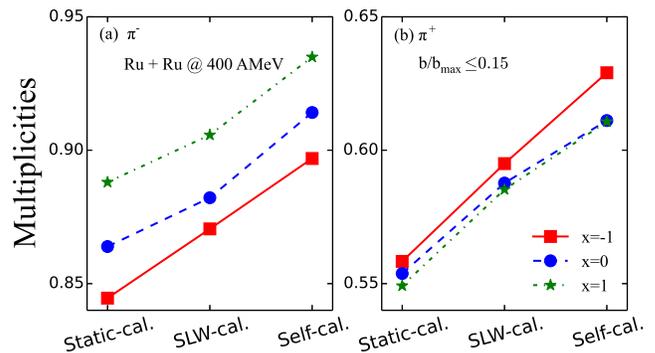

FIG. 3. (Color online) Final multiplicities of $\pi^-$ and $\pi^+$ generated in central $^{96}$Ru+$^{96}$Ru collisions at 400 MeV/nucleon with three calculations scenarios and three symmetry energy settings ranging from the stiff with $x = -1$ to the soft with $x = 1$.

To more clearly show above reasons, we check the relative kinetic energy distribution of nucleons with local densities higher than $\rho_0$ at the reaction maximum com-

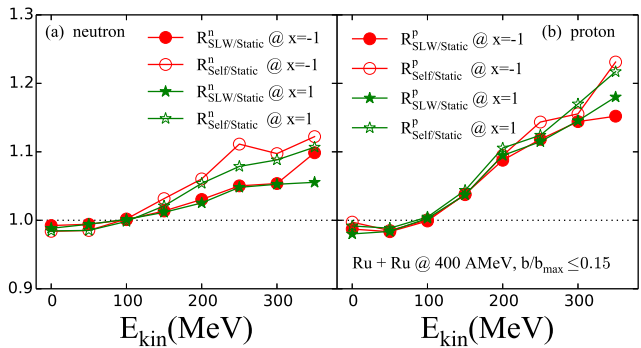

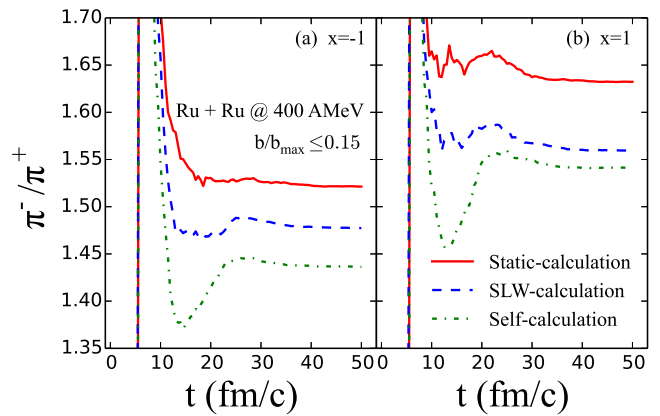

FIG. 4. (Color online) Relative kinetic energy distributions of nucleons with local densities higher than $\rho_0$ at the maximum compress stage in central $^{96}$Ru+$^{96}$Ru collisions at 400 MeV/nucleon.

press stage through the following ratios

$$R^i_{\text{Self/Static}} = \frac{\text{number}(i)_{\text{Self}}}{\text{number}(i)_{\text{Static}}}, \tag{14}$$

$$R^i_{\text{SLW/Static}} = \frac{\text{number}(i)_{\text{SLW}}}{\text{number}(i)_{\text{Static}}}, \tag{15}$$

where $i$ denotes the neutrons or protons, while the symbols "SLW", "Self" and "Static" represent three calculations scenarios for the EM fields, i.e., the SLW formula, the selfconsistent calculation scenario, and the static Coulomb field formula. Shown in Fig. 4 are the relative kinetic energy distributions of nucleons with local densities higher than $\rho_0$ at the maximum compress stage in central $^{96}$Ru+$^{96}$Ru collisions at 400 MeV/nucleon. Obviously, for the energetic nucleons above a certain kinetic energy depending on the beam energy and the reaction system, the values of both $R^n_{\text{Self(SLW)/Staic}}$ and $R^p_{\text{Self(SLW)/Static}}$ are larger than 1, indicating that the weaker anisotropic electric fields and thus the larger cover of high-density region in $Z$ direction indeed increase (decrease) the number of high (low) energy nucleons. Naturally, these increased energetic nucleons are responsible for the increased $\pi^-$ and $\pi^+$ through the $nn$ and $pp$ inelastic channels. Also, that is the same reason, we can also find that the values of $R^n_{\text{Self/Staic}}$ and $R^p_{\text{Self/Static}}$ are larger than that of $R^n_{\text{SLW/Staic}}$ and $R^p_{\text{SLW/Static}}$, respectively. Based on these findings, it is therefore not hard to understand we can see the larger multiplicities of both $\pi^-$ and $\pi^+$ with the selfconsistent calculation scenario than that using the SLW formula, and the latter also leads to the larger multiplicities of both $\pi^-$ and $\pi^+$ than that using the static Coulomb field formula. On the other hand, since the effects of electric fields on protons are direct but indirect on neutrons, as shown also in Fig. 4, the increment of energetic protons is naturally larger than that of energetic nuetrons in both cases using the SLW formula and the selfconsistent scenario for the calculations of EM fields. Consequently, as aforementioned, the resulting effects of electric fields on $\pi^+$ are approximately 2 times larger than that on $\pi^-$ in both cases using the SLW formula and the selfconsistent calculation scenario. This finding naturally leads one to expect the ratio $\pi^-/\pi^+$ to decrease from cases using the static Coulomb field formula to the SLW formula, and this regularity should also hold for cases from using the SLW formula to the selfconsistent calculation scenario.

Shown in Fig. 5 is the evolution of $\pi^-/\pi^+$ ratios generated in central $^{96}$Ru+$^{96}$Ru collisions at 400 MeV/nucleon with three calculations scenarios and two symmetry energy settings ranging from the stiff with $x = -1$ to the soft with $x = 1$. As expected, the $\pi^-/\pi^+$ ratio is significantly smaller in the case using the SLW formula than that using the static Coulomb field formula, and the electric fields with the selfconsistent calculation scenario further decrease the $\pi^-/\pi^+$ ratios compared to that using the SLW formula. Moreover, comparison with the effects of symmetry energy on the $\pi^-/\pi^+$ ratios, effects of the electric fields on them are obviously nonnegligible in cases using either the SLW formula or the selfconsistent calculation scenario. According to these findings, besides the effects of velocity fields, effects of the radiation fields are also very important when using the pion observables to probe the nuclear symmetry energy. Therefore, we conclude that the selfconsistent calculation of the EM fields in HICs should be taken into account although the SLW formula can consider the velocity field effects of EM fields generated in HICs.

FIG. 5. (Color online) Evolution of the $\pi^-/\pi^+$ ratios generated in central $^{96}$Ru+$^{96}$Ru collisions at 400 MeV/nucleon with three calculations scenarios and two symmetry energy settings ranging from the stiff with $x = -1$ to the soft with $x = 1$.

Before ending this part, we examine effects of the electric fields on the double $\pi^-/\pi^+$ ratio of two reactions that is found to have the advantage of maximizing the effects of isovector potential but minimizing that of isoscalar potential as well as electric fields [58, 69]. For this purpose, we show in Fig. 6 the single (upper window) and double (lower window) $\pi^-/\pi^+$ ratios of two isobar reaction systems of $^{96}$Ru+$^{96}$Ru and $^{96}$Zr + $^{96}$Zr at 400 MeV/nucleon. First, since the isospin asymmetry of the reaction $^{96}$Zr +

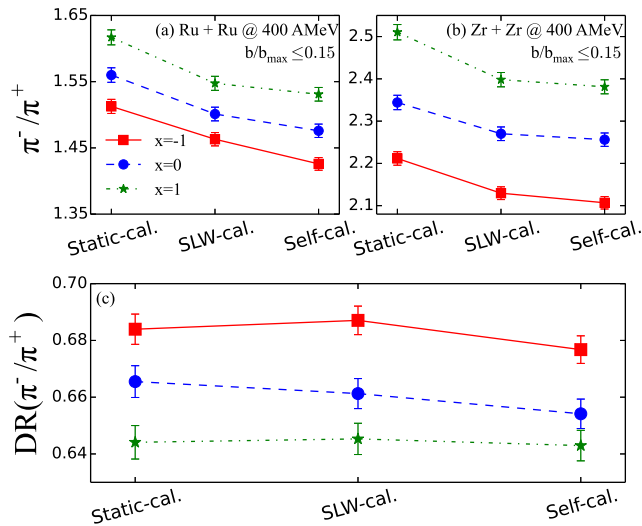

FIG. 6. (Color online) Upper: Final $\pi^-/\pi^+$ ratios generated in two reactions of central $^{96}$Ru+$^{96}$Ru and $^{96}$Zr+$^{96}$Zr collisions at 400 MeV/nucleon. Lower: Double $\pi^-/\pi^+$ ratios (DR($\pi^-/\pi^+$)) of two isobar reaction systems of $^{96}$Ru+$^{96}$Ru and $^{96}$Zr + $^{96}$Zr at 400 MeV/nucleon. Three calculations scenarios and three symmetry energy settings ranging from the stiff with $x=-1$ to the soft with $x=1$ are used.

$^{96}$Zr is larger than that of the reaction $^{96}$Ru+$^{96}$Ru, the $\pi^-/\pi^+$ ratio in the reaction $^{96}$Zr + $^{96}$Zr is more clearly separated by varying the parameter $x$ from -1 to 1 than that in the reaction $^{96}$Ru+$^{96}$Ru. Second, as expected, the double $\pi^-/\pi^+$ ratio of two reactions indeed can effectively eliminate effects of different calculations scenarios for the EM fields but sustain the sensitivity to the nuclear symmetry energy, and thus can still be an effective probe of the nuclear symmetry energy in HICs.

## IV. SUMMARY

In conclusion, we have studied effects of different calculations scenarios for the EM fields generated in HICs on pion observables within a transport model. It is shown that the most used SLW formula is not enough for the electromagnetic field calculation because the absent radiation field in this formula also affects significantly the charged pions as well as their $\pi^-/\pi^+$ ratio. Moreover, the double $\pi^-/\pi^+$ raito of two reactions is found to be less affected by the electromagnetic field calculation scenario and thus can still be an effective probe of the nuclear symmetry energy in HICs. Therefore, according to these findings, we conclude that the selfconsistent calculation of EM fields should be carefully taken into account when using the pion observables especially the multiplicities of charged pions as well as their single $\pi^-/\pi^+$ ratios to probe the nuclear symmetry energy in HICs.

### ACKNOWLEDGMENTS

G. F. Wei would like to thank Profs. Gao-Chan Yong and Yuan Gao for helpful discussions. This work is supported by the National Natural Science Foundation of China under grant Nos.11965008, U1731218, and Guizhou Provincial Science and Technology Foundation under Grant No.[2020]1Y034, and Guizhou Normal University 2019 academic Seedling Cultivation and Innovation Exploration special project, and the PhD-funded project of Guizhou Normal university (Grant No.GZNUD[2018]11).